\documentclass[twocolumn,nofootinbib,aps,showpacs,floats,letterpaper,floatfix,groupedaddress]{revtex4}

\usepackage{dcolumn,epsfig}
\usepackage{amssymb,amsmath}
\usepackage[usenames]{color}
\usepackage{hyperref}

\def\nn{\nonumber}

\def\be{\begin{equation}}
\def\ee{\end{equation}}
\def\beq{\begin{eqnarray}}
\def\eeq{\end{eqnarray}}

\def\IL{\relax{\rm I\kern-.18em L}}

\def\nn{\nonumber}

\def\ba{\begin{eqnarray}}
\def\ea{\end{eqnarray}}

\begin{document}

\title{Gravitational instabilities of superspinars}

\author{Paolo Pani} \email{paolo.pani@ca.infn.it} \affiliation{Dipartimento di
  Fisica, Universit\`a di Cagliari, and INFN sezione di Cagliari, Cittadella
  Universitaria 09042 Monserrato, Italy}

\author{Enrico Barausse} \email{barausse@umd.edu}\affiliation{Maryland Center for Fundamental
    Physics, Department of Physics, University of Maryland, College
    Park, MD 20742, USA}

\author{Emanuele Berti} \email{berti@phy.olemiss.edu} \affiliation{Department
  of Physics and Astronomy, The University of Mississippi, University, MS
  38677-1848, USA}  \affiliation{Theoretical Astrophysics 350-17, California
  Institute of Technology, Pasadena, CA 91125, USA}

\author{Vitor Cardoso} \email{vitor.cardoso@ist.utl.pt}
\affiliation{Centro Multidisciplinar de Astrof\'{\i}sica - CENTRA, Dept. de F\'{\i}sica,
  Instituto Superior T\'ecnico, Av. Rovisco Pais 1, 1049-001 Lisboa, Portugal}\affiliation{Department of Physics and Astronomy, The University of
  Mississippi, University, MS 38677-1848, USA} 

\begin{abstract}
Superspinars are ultracompact objects whose mass $M$ and angular momentum $J$
violate the Kerr bound ($cJ/GM^2>1$). Recent studies analyzed the observable
consequences of gravitational lensing and accretion around superspinars in
astrophysical scenarios. In this paper we investigate the dynamical stability
of superspinars to gravitational perturbations, considering either purely
reflecting or perfectly absorbing boundary conditions at the ``surface'' of
the superspinar. We find that these objects are unstable independently of the
boundary conditions, and that the instability is strongest for relatively
small values of the spin. Also, we give a physical interpretation of the 
various instabilities that we find. Our results (together with the well-known fact that
accretion tends to spin superspinars down) imply that superspinars are very
unlikely astrophysical alternatives to black holes.
\end{abstract}


\maketitle

Superspinars are vacuum solutions of the gravitational field equations whose
mass $M$ and angular momentum $J=aM$ violate the Kerr bound, i.e. $a>M$ (here
and elsewhere in this paper we use geometrical units: $G=c=1$). These
geometries could result from high-energy corrections to Einstein's theory of
gravity, such as those that would be present in string-inspired models
\cite{Gimon:2007ur}. String-inspired corrections may require a modification of
the metric (or some sort of ``excision'') in a small region surrounding the
curvature singularity at the origin, in such a way as to ``dress'' the
singularity.  While stable stars with $a>M$ are in principle allowed in
general relativity\footnote{Typical equations of state usually lead to stars
  with $a/M\lesssim 0.7$ \cite{Cook:1993qr,Berti:2003nb} that can be treated
  within a slow-rotation approximation \cite{Berti:2004ny}.  However, stable,
  differentially rotating polytropic stars with $a/M\approx1.1$ can be
  produced (e.g.) with the Whisky code~\cite{GRS,whisky}. Also, note that the
  Kerr bound can be easily violated by \textit{non-compact} objects such as
  the Earth ($M/R\sim 7\times 10^{-10}$), which has $J/M^2\sim 10^3$.},
superspinars have been proposed as an alternative to black holes (BHs), and
they are therefore imagined to have a compactness comparable to that of
extremal rotating Kerr BHs and to exist in any mass range.  Therefore, the
observation of rapidly spinning ultracompact objects could potentially reveal
or rule out the existence of superspinars.

One argument against the existence of superspinars was put forward in
Ref.~\cite{Cardoso:2008kj}.  There, the authors constructed a toy model for a
superspinar by assuming that the external surface of the superspinar can be
modeled as a perfect mirror, i.e. that the reflection coefficient ${\cal R}=1$
for waves incident on the superspinar. In this case superspinars are
destabilized by superradiant effects, i.e. by the ergoregion instability first
discussed by Friedman, Schutz and Comins \cite{Friedman:1978,Schutz:1978}.
The ergoregion instability occurs on a dynamical timescale, posing a serious
challenge to the existence of these objects in nature. However, a perfectly
reflecting surface may be an unrealistic assumption. In general we would
expect a frequency-dependent reflection coefficient ${\cal R}(\omega)$, and
correspondingly a frequency-dependent transmission coefficient ${\cal
  T}(\omega)=1-{\cal R}(\omega)$. The exact form of ${\cal R}(\omega)$ depends
on the specific model, but unfortunately no exact solutions describing
four-dimensional superspinars are known.

A different instability was recently discussed by Dotti {\it et al.}
\cite{Dotti:2006gc,Dotti:2008yr}. These authors studied perturbations of a
Kerr solution with $a/M>1$ (i.e., unlike Ref.~\cite{Cardoso:2008kj}, they
considered an actual naked singularity). They cast the linearized perturbation
equations in the form of a self-adjoint operator and analyzed the discrete
spectrum of this operator, proving the existence of an infinite number of
unstable modes \cite{Dotti:2008yr}.

Here we generalize the stability analyses of Refs.~\cite{Cardoso:2008kj} and
\cite{Dotti:2008yr} focusing on a superspinar model obtained by considering
the Kerr solution with $a>M$.
Besides extending the study of Ref.~\cite{Cardoso:2008kj}, we also impose an
alternative (and perhaps more physical) prescription for the external surface
of a four-dimensional superspinar. We assume that a perfectly absorbing
surface (a ``stringy horizon'') is created by high-energy effects at some
radius $r=r_0$, and we impose that the reflection coefficient ${\cal
  R}(\omega)\equiv 0$ at that radius.
These purely ingoing boundary conditions at $r=r_0$ are designed to make
superspinars as stable as possible against the ergoregion instability of
Ref.~\cite{Cardoso:2008kj}. This instability occurs because, when the boundary
at $r=r_0$ is purely reflecting, the negative-energy modes which exist in the
ergoregion can only leak to spatial infinity by tunneling through a potential
barrier. Modes propagating outside the ergoregion have positive energies. This
results in the negative energy of the ergoregion modes to decrease
indefinitely, so that their amplitude becomes unbound, triggering an
instability. By imposing purely ingoing boundary conditions at $r=r_0$, we
basically allow the negative energy trapped in the ergoregion to ``fall down a
sink''; this could quench the instability to some extent. A similar quenching
occurs for Kerr BHs with $a\leq M$, the role of the sink being played by the BH
horizon. Clearly, if the reflection coefficient $0<{\cal R}(\omega)<1$ the
quenching would be less efficient. Therefore we conjecture that if
superspinars are unstable when ${\cal R}(\omega)\equiv 0$, it should be
impossible to stabilize them using any other choice of boundary conditions.

In this paper we analyze the stability of superspinars by imposing either
perfectly absorbing (${\cal R}(\omega)\equiv0$) or perfectly reflecting
(${\cal R}(\omega)\equiv1$) boundary conditions at some arbitrary radius
$r=r_0$. We find that, quite independently of $r_0$ and of the chosen boundary
conditions, {\em superspinars are unstable to linearized gravitational
  perturbations.} For purely ingoing boundary conditions the instability is
slightly weaker than in the perfectly reflecting case, but it still occurs on
a dynamical timescale $\tau\sim M$, i.e. $\tau\sim 5\times 10^{-6}$~s for an
object with $M=M_\odot$ and $\tau\sim 5$~s for a supermassive object with
$M\sim10^6 M_\odot$.  We also show that this result is valid for a wide class
of theories of gravity. Our findings undermine several claims made in the
literature that superspinars might be detected because the shadow they cast
due to gravitational lensing \cite{Hioki:2009na,Bambi:2010hf} or their
accretion properties
\cite{Gimon:2007ur,ReinaTreves:1979,Bambi:2009bj,Bambi:2009dw,Takahashi:2010pw,Bambi:2010et}
are different from Kerr BHs with $a<M$. While this is true, superspinars are
plagued by multiple gravitational instabilities, and therefore they are
unlikely to be astrophysically viable BH candidates.

\section{A simple model of superspinar in four dimensions}
Following Gimon and Horava \cite{Gimon:2007ur}, we model a superspinar of mass
$M$ and angular momentum $J=aM$ by the Kerr geometry
\beq
ds_{\rm Kerr}^2&=&-\left(1-\frac{2Mr}{\Sigma}\right)dt^2+\frac{\Sigma}{\Delta}dr^2-\frac{4Mr}{\Sigma}a\sin^2\theta d\phi dt   \nn \\
&+&{\Sigma}d\theta^2+
\left[(r^2+a^2)\sin^2\theta +\frac{2Mr}{\Sigma}a^2\sin^4\theta \right]d\phi^2 \nn\\
\label{kerrmetric}
\eeq
where $\Sigma=r^2+a^2\cos^2\theta$ and $\Delta=r^2+a^2-2M r$.
Unlike Kerr BHs, superspinars have $a/M>1$ and no horizon. Since the domain of
interest is $-\infty<r<+\infty$, the spacetime possesses naked singularities
and closed timelike curves in regions where $g_{\phi\phi}<0$ (see e.g.
\cite{Chandraspecial}).


We study linear perturbations around the Kerr metric (\ref{kerrmetric}).
Using the Kinnersley tetrad and Boyer-Lindquist coordinates, it is possible to
separate the angular and radial variables \cite{Teukolsky:1972my}. Small
perturbations of a spin-$s$ field are then reduced to the radial and angular
master equations
\begin{widetext}
\be
\Delta^{-s}\frac{d}{dr}\left(\Delta^{s+1}\frac{dR_{l
m}}{dr}\right)+ \left[\frac{K^{2}-2is(r-M)K}{\Delta}+4is\omega r -\lambda\right]R_{l m}=0\,,
\label{wave_eq}
\ee
\be
\left[(1-x^2){}_sS_{l m,x} \right]_{,x}+
\left[(a\omega x)^2-2a\omega sx+s+{}_{s}A_{l m}-\frac{(m+sx)^2}{1-x^2}\right]{}_sS_{l m}=0\,,
\label{angularwaveeq}
\ee
\end{widetext}
where $x\equiv\cos\theta$, $K=(r^2+a^2)\omega-am$ and the separation constants
$\lambda$ and ${}_s A_{l m}$ are related by
\be
\lambda \equiv {}_s A_{l m}+a^2\omega^2-2am\omega\,.
\label{sAlm}
\ee
The equations above describe scalar, electromagnetic and gravitational
perturbations when $s=0,\,\pm1,\,\pm2$ respectively. 
The oscillation frequencies of the modes can be found from the canonical form
of Eq.~(\ref{wave_eq}). Switching to a ``tortoise coordinate'' $r_*$ defined
by the condition $dr_*/dr=(r^2+a^2)/\Delta$, we get
\be
\frac{d^2Y}{dr_*^2}+VY=0\,,
\label{teu canonical}
\ee
where
\beq
Y&=&\Delta^{s/2}(r^2+a^2)^{1/2}R\,,\nn\\
V&=&\frac{K^2-2is(r-M)K+\Delta(4ir\omega
s-\lambda)}{(r^2+a^2)^{2}}-G^2-\frac{dG}{dr_*}\,,\nn
\eeq
and $G=s(r-M)/(r^2+a^2)+r\Delta(r^2+a^2)^{-2}$. The eigenvalues ${}_sA_{lm}$
in Eq.~(\ref{sAlm}) can be expanded in a power series in the parameter
$a\omega$ as \cite{Berti:2005gp}
\be {}_sA_{lm}=\sum_{n=0}f^{(n)}_{slm}(a\omega)^{n}\,.
\label{rel:eigexpans} \ee
The absence of ingoing waves at infinity implies \cite{Teukolsky:1974yv}
\be
Y\sim r^{-s}e^{i\omega r_*}\,,\qquad\qquad r\to\infty\,.\label{asymp sol}
\ee

The boundary conditions at $r=r_0$ are crucial. Ref.~\cite{Cardoso:2008kj}
assumed a perfect mirror at $r=r_0$, i.e. $Y(r_0)=0$. If instead we assume the
existence of some ``stringy horizon'' at $r_0$, we must impose purely ingoing
waves as $r\to r_0$.  Since for $a>M$ the potential $V$ is regular at any
$r=r_0$ (including also $r_0/M=0$), we can write
\be\label{Vseries}
V(r)\sim V(r_0)+{\cal O}(r-r_0)\,.
\ee
By expanding Eq.~\eqref{teu canonical} in series around $r=r_0$ we find that
the general solution is a superposition of ingoing and outgoing waves:
\be
Y\sim Ae^{-i k r_*}+Be^{i k r_*}+O(r_*-r_*(r_0))^3\,,
\qquad k^2=V(r_0)\,,\label{BCingoing} 
\ee
where the sign of $k$ is chosen to recover the well-known boundary condition
for a wave-function in an extreme Kerr background ($a\to M$ and $r_0\to M$):
$k=\omega-m\Omega$, where $\Omega=1/(2M)$ is the angular velocity of an
extreme Kerr black hole. Purely ingoing boundary conditions at the stringy
horizon imply $B=0$ in Eq.~\eqref{BCingoing} or, equivalently,
\be
\frac{dY}{dr_*}=-ikY\,,\qquad r\to r_0\,.
\label{BCstringy}
\ee
This is the condition we impose in our numerical code.

For each $\omega$, we integrate Eq.~(\ref{teu canonical}) numerically
inward, starting at some large radius (typically $r_\infty = 400M$) where we
impose the asymptotic behavior (\ref{asymp sol}). Our results are robust to
variations of $r_{\infty}$ in a reasonable range.
We stop the numerical integration at $r=r_0$, where the value of the field
$Y(\omega,r_0)$ is extracted. Finally, we repeat the integration for different
values of $\omega$ until the desired boundary condition (either $Y(\omega,r_0)=0$ or Eq.~\eqref{BCstringy}) at $r=r_0$ is
satisfied, typically to within an accuracy of $10^{-10}$.

In our numerical computations we make use of the series expansion
(\ref{rel:eigexpans}), truncated at fourth order. When $|a\omega|<1$, the
series expansion is a very good approximation of the exact
eigenvalues. However, in some cases (i.e. when $|a\omega|\gtrsim1$), instead
of the series expansion we have used exact numerical values of ${}_s A_{lm}$
obtained by solving Eq.~(\ref{angularwaveeq}) with the continued fraction
method \cite{Berti:2009kk}.

We focus on the most relevant gravitational perturbations, described by the
Teukolsky equation with $s=2$.  To compute unstable modes we also make use of
the symmetry \cite{Leaver:1985ax,Berti:2009kk}
\be
m\to-m\,,\quad\omega\to-\omega^*\,,\quad {}_sA_{lm} \to{}_sA_{l-m}^*\,.\label{symmetry}
\ee
In practice, this symmetry means that modes with azimuthal number $-m$ can be
obtained from those with azimuthal number $m$ by changing the sign of the real
part of the frequency. Therefore we focus on modes with ${\rm Re}[\omega]=\omega_R>0$ only.

\begin{figure*}[thb]
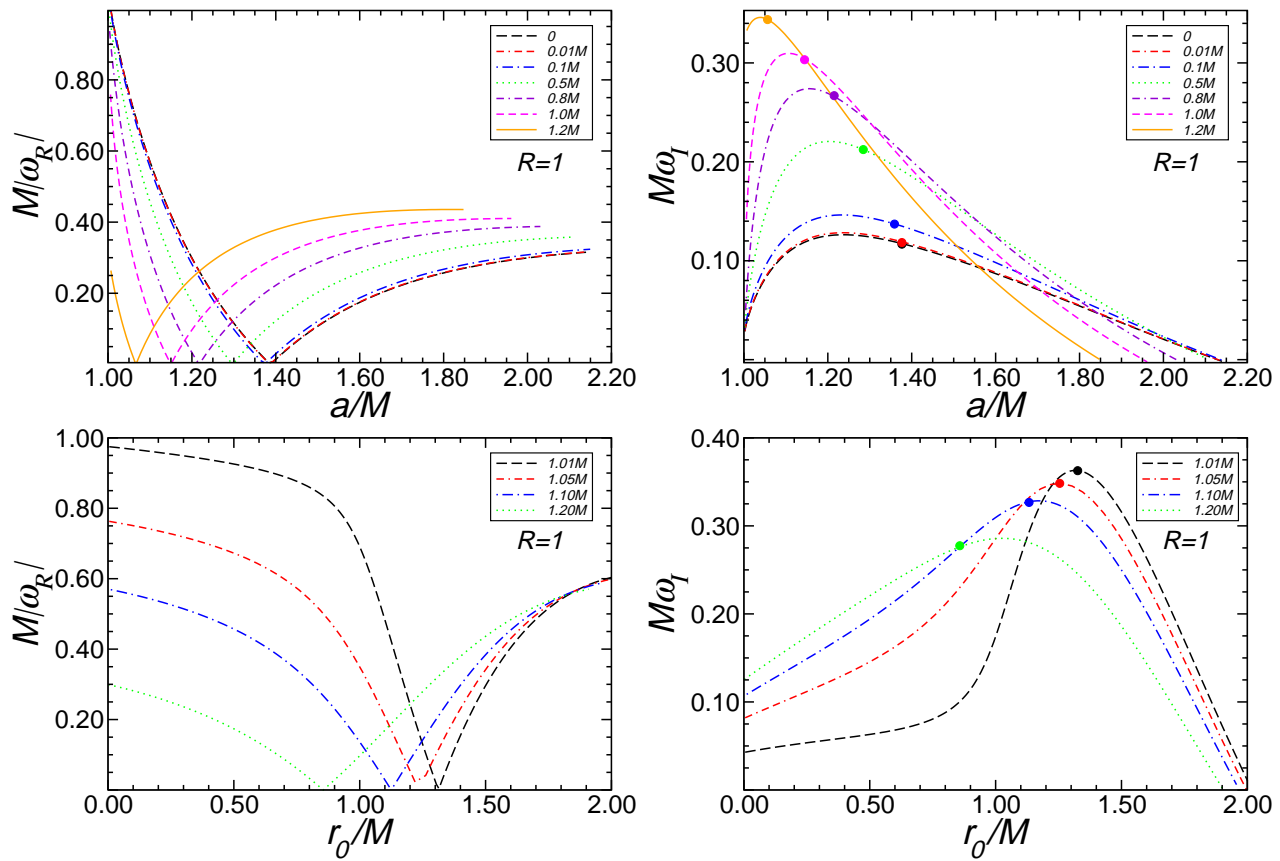

\begin{center}
\begin{tabular}{cc}
\includegraphics[scale=0.31,clip=true]{./plots/wR_VS_a_bis.eps}&
\includegraphics[scale=0.31,clip=true]{./plots/wI_VS_a_bis.eps}\\
\includegraphics[scale=0.31,clip=true]{./plots/wR_VS_r0.eps}&
\includegraphics[scale=0.31,clip=true]{./plots/wI_VS_r0.eps}
\end{tabular}
\end{center}
\caption{\label{fig:wVSa} Top: Real (left) and imaginary part (right) of
  unstable gravitational modes of a superspinar as a function of the spin
  parameter, $a/M$, for $l=m=2$ and several fixed values of $r_0$. Bottom:
  Real (left) and imaginary part (right) of unstable gravitational modes of a
  superspinar as a function of the mirror location, $r_0/M$, for $l=m=2$ and
  different fixed values of the spin parameter. Large dots indicate purely
  imaginary modes.}
\end{figure*}
%

\section{Perfect mirror }
\subsection{Unstable modes with $l=m=2$}

Let us start by reviewing and extending the results of
Ref.~\cite{Cardoso:2008kj}, which first found that superspinars with a perfectly reflecting
surface are unstable due to the ergoregion instability. In the top panel of 
Fig.~\ref{fig:wVSa} we show unstable frequencies for $s=l=m=2$ as
a function of the spin parameter $a/M$ for selected values of $r_0/M$. We see
that the instability (signalled by a positive imaginary part $\omega_I$ for the frequency) 
is always strong, i.e. it always occurs on a short
timescale $\tau\equiv 1/\omega_I\sim10M\sim 5\times 10^{-5}(M/M_\odot)$~s, at
least when $a\lesssim2.2M$. It is interesting to note that the instability is
also effective for $r_0=M$ and $a=M+\epsilon$, i.e. for an object as compact
as an extremal Kerr BH with a rotation parameter that only slightly violates
the Kerr bound. This is also illustrated in Table \ref{tab:wVSa}.

\begin{widetext}
\begin{center}
\begin{table}[th]
\caption{\label{tab:wVSa} Unstable gravitational ($s=2$) frequencies with
  $l=m=2$ for a superspinar with a perfect reflecting surface (${\cal R}=1$)
  and with a ``stringy event horizon'' (${\cal R}=0$) at $r=r_0$. All modes in
  this table have been computed using numerical values of ${}_sA_{lm}$
  obtained via the continued fraction method \cite{Berti:2009kk}.}
\begin{tabular}{c||ccc|ccc}
%
\hline \multicolumn{1}{c}{} & \multicolumn{3}{c}{ $(\omega_R M\,,\omega_I M),\,\,\,{\cal R}=1$} &\multicolumn{3}{c}{ $(\omega_R M\,,\omega_I M),\,\,\,{\cal R}=0$}\\
\hline
$r_0/M$  & $a=1.1M$            &$a=1.01M$           &$a=1.001M$          & $a=1.1M$            &$a=1.01M$         	&$a=1.001M$\\
$0.01$   & $(0.5690\,,0.1085)$ &$(0.9744,,0.0431)$  &$(0.9810\,,0.0097)$ &$(0.5002\,,0.0173)$  &$(0.9498\,,0.0062)$	&$(1.0286\,,0.0033)$    \\
$0.1$    & $(0.5548\,,0.1237)$ &$(0.9673\,,0.0475)$	&$(0.9794\,,0.0110)$ &$(0.4878\,,0.0260)$  &$(0.9435\,,0.0093)$	&$(1.0252\,,0.0048)$   \\
$0.5$    & $(0.4571\,,0.1941)$ &$(0.9256\,,0.0631)$	&$(0.9688\,,0.0155)$ &$(0.3959\,,0.0719)$  &$(0.9016\,,0.0237)$	&$(1.0052.\,,0.0091)$   \\
$0.8$    & $(0.3081\,,0.2617)$ &$(0.8598\,,0.0878)$	&$(0.9507\,,0.0202)$ &$(0.2537\,,0.1053)$  &$(0.8298\,,0.0376)$	&$(0.9793\,,0.0095)$   \\
$1$      & $(0.1364\,,0.3095)$ &$(0.6910\,,0.1742)$	&$(0.9003\,,0.0640)$ &$(0.0916\,,0.1219)$  &$(0.6530\,,0.0821)$	&$(0.8853\,,0.0313)$   \\
$1.1$    & $(0.0286\,,0.3248)$ &$(0.4831\,,0.2655)$	&$(0.6071\,,0.2207)$ &$(-0.0078\,,0.1233)$ &$(0.4377\,,0.1230)$	&$(0.5696\,,0.1064)$   \\
\hline \hline
\end{tabular}
\end{table}
\end{center}
\end{widetext}

From Fig.~\ref{fig:wVSa} and Table~\ref{tab:wVSa}, it is clear that when
$r_0>M$ the imaginary part does not vanish as $a\to M$. This is in agreement
with our expectations, since when $r_0>r_H$ the ``BH bomb'' instability
\cite{Press:1972,Cardoso:2008kj} occurs even when $a<M$.

The dependence of the eigenfrequencies on the mirror location is also shown in
the bottom panel of Fig.~\ref{fig:wVSa} for different values of the spin
parameter. The imaginary part of the frequency is positive (i.e. the object
is unstable) for a wide range of parameters. For any value of $a/M$ in the
bottom panel of Fig.~\ref{fig:wVSa} the instability is strongest when
$r_0/M\sim1$, and is effective also in the limit $r_0/M\ll 1$ 
(although in this regime high-energy corrections to the background metric could be
relevant).

Overall, Fig.~\ref{fig:wVSa} shows that the strongest instability occurs
roughly when $a/M \sim1.1$.  For larger values of the spin the imaginary part
decreases and eventually it vanishes (causing the instability to disappear)
for a critical value of $a/M$ which depends on $r_0$.  At first sight, this
result seems in contrast with the superradiant nature of the instability, as
one may naively think that the instability should become stronger for large
spins.

\begin{figure}[htb]
\begin{center}
\begin{tabular}{c}
\includegraphics[scale=0.31,clip=true]{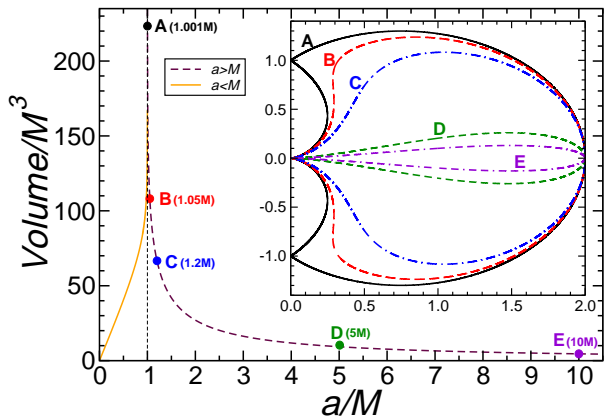}
\end{tabular}
\end{center}
\caption{\label{fig:ergoregion} Proper volume of the ergoregion as a function
  of the spin $a/M$. The volume increases monotonically when $a<M$, is
  infinite at $a=M$ and decreases monotonically when $a>M$. The proper volumes
  for $a\sim2M$ and $a\sim0.3M$ are roughly the same. In the inset we show the
  azimuthal section of the ergoregion for selected values of the spin. These
  spins are marked by filled circles and capital Latin letters in the main
  plot; their numerical value is indicated in parentheses in the figure. In
  the limit $a/M\to\infty$ the ergoregion becomes so oblate that its proper
  volume shrinks to zero.}
\end{figure}

In Fig.~\ref{fig:ergoregion} we show that this expectation is not justified by
plotting the proper volume of the ergoregion as a function of $a/M$. The
proper volume can be computed via
\be
V=4\pi\int_{\theta_i}^{\pi/2} d\theta\int_{r_i}^{r_f}dr\sqrt{g_{rr}g_{\theta\theta}g_{\varphi\varphi}}\,,\label{prop_vol}
\ee
where we have considered a constant time slice, the metric elements are taken
from Eq.~(\ref{kerrmetric}), we have exploited the reflection symmetry of the
Kerr metric, and we have already integrated out the $\varphi$ dependence. For
$a<M$ the ergoregion extends between the outer Kerr horizon at
$r_H=M+\sqrt{M^2-a^2}$ and the ``outer ergosphere radius'' at
$r_{e+}(\theta)=M+\sqrt{M^2-a^2\cos^2\theta}$. In this case we set $r_i=r_H$,
$r_f=r_{e+}$, and $\theta_i=0$ in the integral above.  A straightforward
calculation shows that in this case the proper volume increases monotonically
with $a/M$, eventually diverging\footnote{This is because when $a=M$,
  $g_{rr}\sim1/(r-M)^2$ near the horizon $r=M$: this causes the
  integral~\eqref{prop_vol} to diverge logarithmically.} for $a=M$.  However,
when $a>M$, the ergoregion extends between the inner ergosphere at
$r_{e-}(\theta)=M-\sqrt{M^2-a^2\cos^2\theta}$ and the outer ergosphere
$r_{e+}(\theta)$.  Therefore we set $r_i=r_{e-}$, $r_f=r_{e+}$, and
$\theta_i=\arccos(M/a)$ in the integral (\ref{prop_vol}). In this case, the
proper volume of the ergosphere monotonically {\it decreases} with $a/M$. In
the inset of Fig.~\ref{fig:ergoregion} we plot an azimuthal section of the
ergoregion for selected values of the spin parameter. The proper volume of the
ergoregion vanishes as $a/M\to \infty$ because the ergoregion becomes more and
more oblate (in the equatorial direction) as the spin increases. As $a/M\to
\infty$ the proper volume shrinks to zero and the ergoregion instability for
modes with $l=m=2$ becomes harmless. In Section~\ref{phys_origin}, however, we
will see that this suppression of the ergoregion instability is less effective
for modes with $l=m\gg2$, which are more concentrated in the equatorial region
and which make superspinars unstable even for larger values of the spin.

\begin{figure*}[htb]
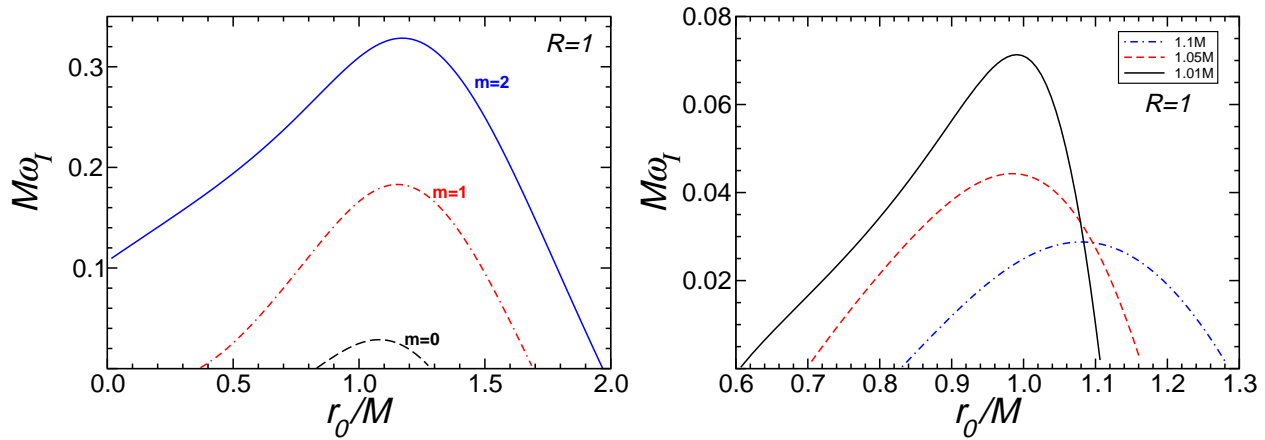

\begin{center}
\begin{tabular}{cc}
\includegraphics[scale=0.31,clip=true]{./plots/wI_VS_r0_m012.eps}&
\includegraphics[scale=0.31,clip=true]{./plots/wI_VS_r0_m0.eps}
\end{tabular}
\end{center}
\caption{\label{fig:m012} Left: Imaginary part of unstable gravitational modes
  of a superspinar as a function of the mirror location, $r_0/M$, for
  $a=1.1M$, $l=2$ and $m=0,\,1,\,2$. Right: Imaginary part of unstable
  gravitational modes of a superspinar as a function of the mirror location,
  $r_0/M$, for $l=2$, $m=0$ and several values of the spin parameter, $a$.}
\end{figure*}
\subsection{Unstable modes with $m=0$}
Superradiance due to an ergoregion is not the only mechanism driving
instabilities in superspinars. Unstable modes also exist for $m=0$, when the
condition for superradiance $\omega<m\Omega=0$ cannot be fulfilled. This is
shown in the left panel of Fig.~\ref{fig:m012}, where we show different
gravitationally unstable modes for $l=2$ and $m=0,1,2$.

We see that the unstable mode with $m=2$ exists in the range $0\leq r_0/M\leq
2$, i.e. out to the outer location of the ergoregion in the equatorial
plane. Unstable modes with $m=1$ and $m=0$ exist only in a more limited range
around $r_0\sim M$. Modes with larger values of $m(\leq l)$ drive stronger
instabilities, but the instability for modes with $m=0$ is also important,
because it occurs on a dynamical timescale $\tau=1/\omega_I \sim M$ for a wide
range of parameters. In the right panel of Fig.~\ref{fig:m012} we show some
unstable modes with $m=0$ for different values of the spin parameter.

\begin{figure*}[htb]
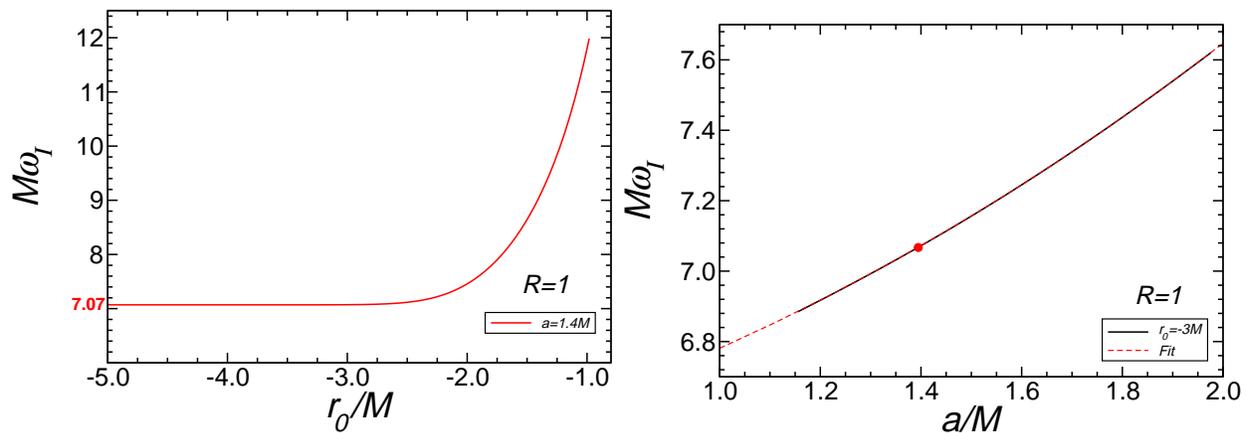

\begin{center}
\begin{tabular}{cc}
\includegraphics[scale=0.31,clip=true]{./plots/wI_VS_r0_dotti.eps}&
\includegraphics[scale=0.31,clip=true]{./plots/wI_VS_a_dotti3.eps}
\end{tabular}
\end{center}
\caption{\label{fig:dotti} Left: Purely imaginary unstable mode as a function
  of the mirror location, $r_0/M<0$, for $a=1.4M$, $s=l=2$ and $m=0$. In the
  limit $r_0\to-\infty$, $M\omega_I\sim7.07$, which perfectly agrees with
  results in Ref.~\cite{Dotti:2008yr}. Right: Purely imaginary unstable mode
  as a function of the spin, $a/M$, for $r_0=-3M$, $s=l=2$, $m=0$. Numerical
  results (black straight line) are consistent with the quadratic fit of
  Eq.~(\ref{fitdotti}) (red dashed line). The dot marks the case considered in
  the left panel.}
\end{figure*}

Unstable modes with $m=0$ have been recently found by Dotti and Gleiser
\cite{Dotti:2006gc,Dotti:2008yr}. By imposing regularity conditions at
$r=-\infty$ these authors found (an infinite number of) unstable purely
imaginary
modes when $a>M$. The superspinar model we are discussing reduces to the
spacetime considered in Refs.~\cite{Dotti:2006gc,Dotti:2008yr} when
$r_0\to-\infty$ and ${\cal R}=1$.  We carried out a search of these purely
imaginary unstable modes, and our results agree very well with those of
Ref.~\cite{Dotti:2008yr} in this limit. For illustration, in the left panel of
Fig.~\ref{fig:dotti} we show that the frequency of the $m=0$ purely imaginary
mode for $a=1.4 M$ matches the result of Ref.~\cite{Dotti:2008yr} for
$r_0\to-\infty$. The figure shows that the frequency of these modes settles to
its asymptotic value when $r_0\lesssim-3M$, so it makes sense to fit $M
\omega_I$ for these $m=0$ purely imaginary modes as a function of $a$, setting
$r_0=-3 M$.  A comparison between the numerical results and the polynomial fit
\be\label{fitdotti} 
M\omega_I=6.375 + 0.177 a/M + 0.230 (a/M)^2 
\ee 
is presented in the right panel of Fig.~\ref{fig:dotti}. In the range $1.15\lesssim a/M<2$ the fit is accurate
to within $0.003\%$ and suggests that the
imaginary part of the frequency (and therefore the ``strength'' of the
instability) grows approximately as a quadratic function of the spin.  We
stress that these results have been obtained using the asymptotic expansion of
the angular spheroidal eigenvalues \cite{Berti:2005gp}
\be
{}_sA_{lm}\sim(2l-\eta+1)|a\omega|-s(s+1)+{\cal O}(|a\omega|^0)\,,
\ee
where $\eta=2\max\left(|m|,|s|\right)$. Because for these modes $|a\omega|\gtrsim 5$, this expansion
is a good approximation of the
numerical eigenvalues computed by the continued fraction method
\cite{Berti:2005gp}. 
%

\section{Absorbing boundary conditions (horizon-like surface at $r=r_0$)\label{sec:R=0}}

From the results discussed in the previous section we conclude that a
dynamical instability is almost unavoidable in a broad region of the parameter
space if the surface of the superspinar is perfectly reflecting. The
instability is present even in what would naively seem the most
phenomenologically viable case, i.e. when $r_0\sim M$ and $a=M+\epsilon$. One
could argue that a perfectly reflecting surface maximizes the efficiency of
the ergoregion instability because negative-energy modes, which are
potentially dangerous, cannot be absorbed, and that this might not happen for
different boundary conditions. In fact, Kerr BHs are stable because (despite
superradiant scattering) the negative-energy modes can flow down the horizon.
Therefore we expect ingoing boundary conditions (${\cal R}=0$) at $r=r_0$ to
represent the worst possible situation for the ergoregion instability to
develop. If we find an instability even in this case, the superspinars
described by our simple model are doomed to be unstable.  This choice also
seems more physically motivated than the perfectly reflecting boundary
conditions, because $r_0$ might be the location of an event horizon formed by
string-inspired modifications of gravity at high curvatures.
We consider both modes with $l=m=2$ (which we expect to be affected by the
ergoregion instability) as well as modes with $m=0$, which we found to be
unstable in the perfectly reflecting case.

\begin{figure*}[htb]
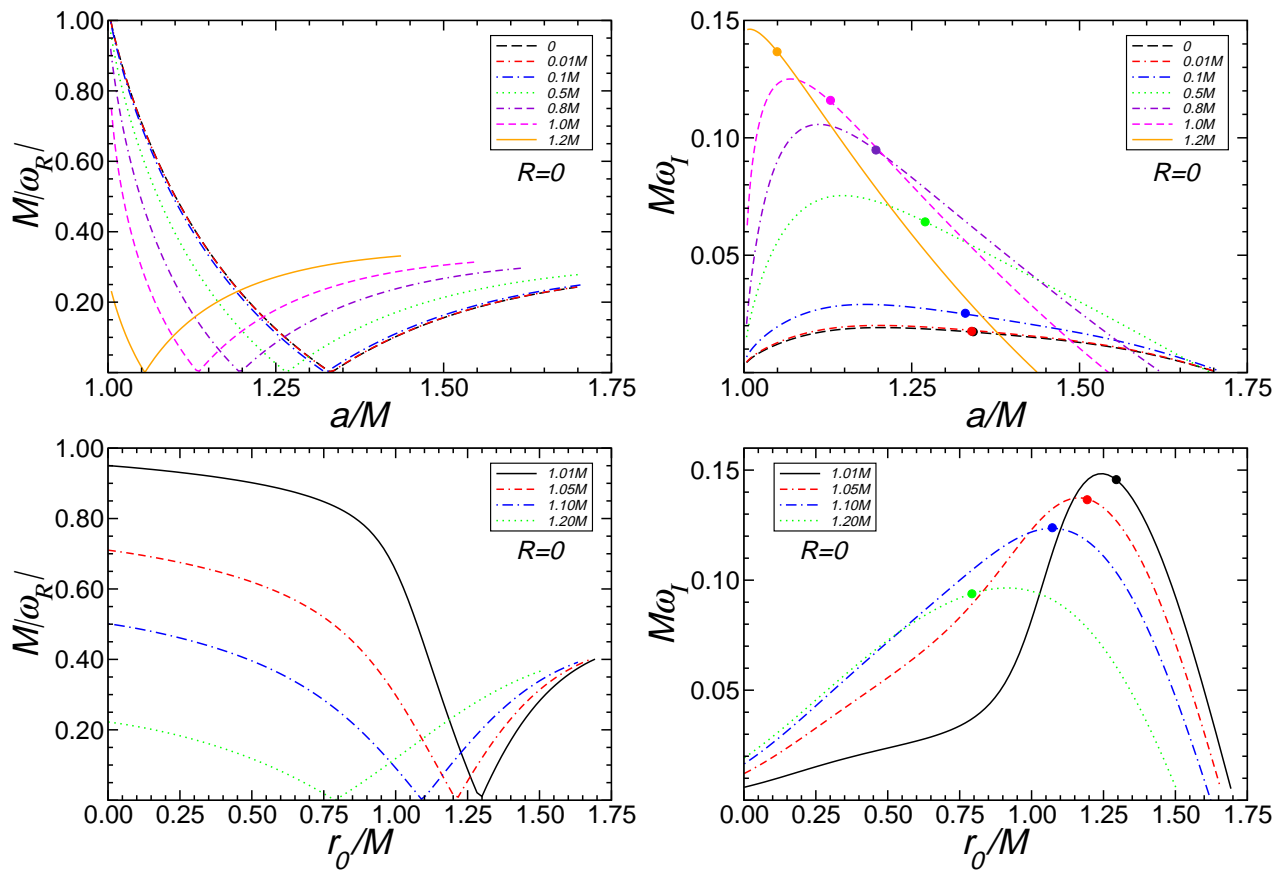

\begin{center}
\begin{tabular}{cc}
\includegraphics[scale=0.31,clip=true]{./plots/wR_VS_a_R0_bis.eps}&
\includegraphics[scale=0.31,clip=true]{./plots/wI_VS_a_R0_bis.eps}\\
\includegraphics[scale=0.31,clip=true]{./plots/wR_VS_r0_R0.eps}&
\includegraphics[scale=0.31,clip=true]{./plots/wI_VS_r0_R0.eps}
\end{tabular}
\end{center}
\caption{\label{fig:wVSa_R0} Top: Real (left) and imaginary part (right) of
  unstable gravitational modes of a superspinar as a function of the spin
  parameter, $a/M$, for $l=m=2$ and several fixed values of the horizon
  location $r_0/M$. Bottom: Real (left) and imaginary part (right) of unstable
  gravitational modes of a superspinar as a function of the horizon
  location, for $l=m=2$ and fixed values of the spin parameter. Large dots
  indicate purely imaginary modes.
}
\end{figure*}

The punchline of this section is that unstable modes exist even when we impose
ingoing boundary conditions. Qualitatively, the results are the same as those
obtained by imposing perfect reflection at the surface of the superspinar. The
instability is slightly weaker than in the previous case, but it is again
unavoidable in a wide region of parameter space.

Gravitationally unstable modes with $l=m=2$ and ${\cal R}=0$ are listed in Table
\ref{tab:wVSa} and shown in Fig.~\ref{fig:wVSa_R0} (to be compared with
Fig.~\ref{fig:wVSa}).  Typically the imaginary part of the unstable modes when
${\cal R}=0$ is only one order of magnitude smaller than that obtained
imposing ${\cal R}=1$, which causes the instability to disappear at slightly smaller spins, \textit{i.e.} when $a/M\gtrsim1.75$.
However, as already mentioned, in the next section we will present 
evidence that higher-$l$ modes are unstable for larger values of the spin and show that
our results are sufficient to rule out superspinars as
astrophysically viable alternatives to Kerr BHs.

Also, we stress that perfectly absorbing ``stringy horizons'' can only be created by high-energy
effects taking place beyond the range of validity of general relativity. From
a phenomenological point of view, the region that should be modified by these
high-energy corrections is close to the curvature singularity of the Kerr
metric ($r_0\ll M$).  Unstable modes generically exists for $a=M+\epsilon$,
even in the limit $r_0/M\to 0$. Moreover, results are smooth in the limit
$r_0/M\to 0$, which means that in the region spanned by our calculations
curvature singularities do not affect our conclusions.

\begin{figure}[htb]
\begin{center}
\begin{tabular}{c}
\includegraphics[scale=0.31,clip=true]{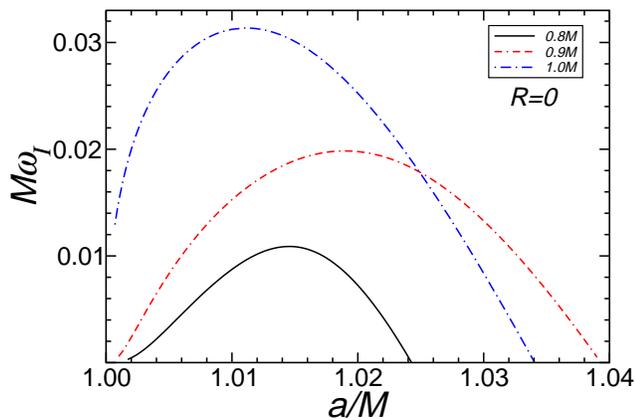}
\end{tabular}
\end{center}
\caption{\label{fig:m0_R0} Imaginary part of unstable gravitational
  modes of a superspinar as the spin parameter, $a$, for $l=2$, $m=0$ and
  several values of the horizon location, $r_0$. 
}
\end{figure}

Finally, Fig.~\ref{fig:m0_R0} (to be compared with Fig.~\ref{fig:m012}) shows that unstable modes with $m=0$ are still present
when we impose ${\cal R}=0$ at $r=r_0$.

\begin{figure}[htb]
\begin{center}
\begin{tabular}{c}
\includegraphics[scale=0.31,clip=true]{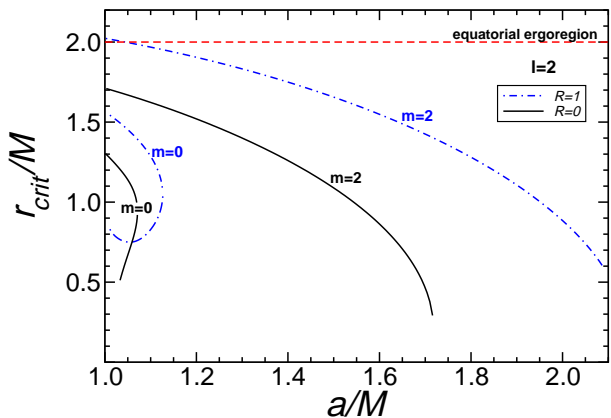}
\end{tabular}
\end{center}
\caption{\label{fig:rcrit} Critical radius $r_{\rm crit}/M$ as a function of
  $a/M$ for $s=l=2$ and $m=0,\,2$. We impose both ${\cal R}=0$ and ${\cal
    R}=1$ at $r=r_0$. For $m=2$ an instability occurs when $r_0<r_{\rm
    crit}$. For $m=0$ an instability occurs in the region delimited by the
  curves on the left. When $l=m\gg1$ an instability is expected to occur also
  when $a\gtrsim2.2M$ in the ${\cal R}=1$ case, and when $a\gtrsim1.75M$
 in the ${\cal R}=0$ case (cf. Fig.~\ref{fig:l2_l5}). Although not shown, the
  region where $r_0/M<0$ is expected to be unstable for any value of $a>M$.}
\end{figure}
%

\section{Modes with $l=m\gg1$ and physical origin of the instability}
\label{phys_origin}
%
We have seen that when we impose perfectly reflecting boundary conditions
(${\cal R}=1$) superspinars are plagued by several instabilities. Moreover,
these instabilities are still present when we impose ${\cal R}=0$, i.e. when
we consider a ``stringy horizon'' at $r=r_0$, which would be expected to
quench the instabilities.  Here we analyze in more detail how these different
instabilities arise.  We focus first on the case $r_0/M>0$. For ${\cal R}=1$ from
Figs.~\ref{fig:wVSa}~and~\ref{fig:m012}  (and analogously for ${\cal R}=0$ from Figs.~\ref{fig:wVSa_R0}~and~\ref{fig:m0_R0}) we
see that, for both $m=0$ and $m=2$, the imaginary part vanishes at some
critical radius: $\omega_I(r_{\rm crit})=0$.  By using a root-finding routine
we can solve for the critical radius as a function of the spin parameter
$a/M$.  The results are shown in Fig.~\ref{fig:rcrit}. For $m=2$ and ${\cal
  R}=1$ the instability occurs in the region below the dot-dashed blue line
extending from $a\sim M$ up to $a/M\sim2.2$. The dashed horizontal line marks
the location of the outer ergoregion on the equatorial plane ($r=2M$). As
$a/M\to 1$ the critical radius roughly coincides with the location of the
ergoregion, and it decreases monotonically for larger rotation. The situation
is similar for $m=2$ and ${\cal R}=0$ (solid black line extending from $a\sim
M$ up to $a/M\sim1.75$), with the instability disappearing earlier (at
$a/M\sim1.75$) because the ingoing boundary conditions allow negative energy
modes to flow down the stringy horizon. This plot confirms our qualitative
understanding of the instability: as shown in Fig.~\ref{fig:ergoregion}, in
the limit $a/M\to \infty$ the proper volume of the ergoregion vanishes and
superradiance cannot destabilize the modes at arbitrarily large spins.

\begin{figure}[htb]
\begin{center}
\begin{tabular}{c}
\includegraphics[scale=0.31,clip=true]{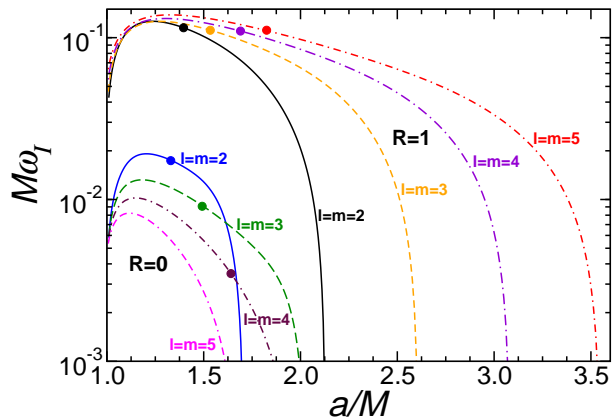}
\end{tabular}
\end{center}
\caption{\label{fig:l2_l5} Imaginary part of the
  fundamental unstable mode as a function of the spin $a/M$ for a superspinar
  with $r_0=0$. Upper curves refer to ${\cal R}=1$ with $l=m=2,3,4,5$. Lower curves refer to ${\cal R}=0$ with $l=m=2,3,4,5$. Dots indicate purely imaginary modes.}
\end{figure}

It would therefore seem that superspinars with $a>2.2M$ might be stable, for
any value of $r_0$ and even in the most restrictive case with ${\cal R}=1$. We
argue, however, that this is not true if we consider modes with $l=m>2$.  In
fact the angular distribution of modes with higher $l=m$ is more concentrated
around the equatorial plane. Higher-$l$ modes become more effective at
destabilizing the superspinar when the ergoregion is oblate, i.e. when $a\gg
M$ (see again the inset of Fig.~\ref{fig:ergoregion}). This intuitive
understanding is confirmed by Fig.~\ref{fig:l2_l5}. There we plot the
imaginary part of the fundamental unstable modes for $l=m=2$ and $l=m>2$
(setting $r_0=0$), both for ${\cal R}=1$ and ${\cal R}=0$. 

In the case ${\cal R}=1$, unstable modes with higher $l=m$ exist for larger
values of the spin. For example the $l=m=5$ mode becomes stable when
$a/M\gtrsim3.6$, while the $l=m=2$ mode becomes stable when $a/M\gtrsim2.2$,
as previously discussed. Modes with $l=m\gg1$ are generally difficult to
compute with our code. Our results suggest that, for any fixed value of
$a/M\gg1$, there are always unstable modes as long as $l=m$ is sufficiently
large.  We stress that these results are in contrast with the case of Kerr
BHs, where the superradiant amplification is always stronger for $l=m=2$
\cite{Teukolsky:1974yv}.  Results are qualitatively similar for ${\cal R}=0$,
in which case the $l=m=3$ instability disappears when $a/M\gtrsim2$, while the
$l=m=2$ instability disappears when $a/M\gtrsim1.75$, as previously
discussed. However in this case the $l=m=4$ instability is weaker than the
$l=m=3$ instability, and the $l=m=5$ instability disappears for smaller values
of $a/M$ than in the $l=m=2$ case. This is because the stringy event horizon
quenches the instability of higher-$l$ modes, similarly to the horizon of a
Kerr BH.

More in general, the ergoregion instability that we have found here at the
linear level can be related to simple kinematical properties of the Kerr
spacetime. In fact, as we show in Appendix \ref{app:geod_Kerr}, Kerr
spacetimes with $a>M$ admit \textit{stable} non-equatorial null circular
orbits with \textit{negative} energy. These orbits exist at any radius
$r<M$. The very existence of these orbits is enough to prove that the
spacetime is plagued by the ergoregion instability, \emph{provided} that purely
reflecting boundary conditions (${\cal R}=1$) are imposed at $r_0<M$. This is
because null orbits are the geometric-optics limit of gravitational
perturbations, as can easily be seen by expanding their propagation equation
in powers of $1/\lambda$ ($\lambda$ being the wavelength of the
perturbation). Therefore, the existence of stable null circular orbits with
negative energies implies the existence of short-wavelength modes with negative energies. Under perfectly reflecting boundary conditions, these
modes can only leak to infinity by tunneling through the potential
barrier. However, because particles outside the ergoregion must have positive
energies, this leak makes the energy of the perturbations inside the
ergoregion more and more negative. As a result, their amplitude grows without
bound, thus revealing the instability of the spacetime.
For these reasons we expect modes with $l=m\gg1$ to be unstable for
arbitrarily large values of $a/M$, at least for ${\cal R}=1$.  This
expectation is consistent with the general theorem of
Ref.~\cite{Friedman:1978}, which states that any spacetime possessing an
ergoregion, but not an event horizon, is vulnerable to the ergoregion
instability. As shown in Fig.~\ref{fig:l2_l5}, this expectation is not always
justified in the less efficient case with ${\cal R}=0$.

The existence of stable non-equatorial null circular orbits with negative
energy clarifies why superspinars are unstable even under perfectly absorbing
boundary conditions (i.e., in the presence of a stringy horizon), while for
Kerr BHs with $a\leq M$ the presence of the horizon kills the ergoregion
instability. For superspinars, the effective potential for gravitational
perturbations presents a minimum at small radii (corresponding, in the eikonal
limit, to the location of a negative-energy stable non-equatorial null
circular orbit), and then rises as $r/M\sim0$. Therefore, the ergoregion modes
need to tunnel through a potential barrier to fall into the ``stringy''
horizon. The stability of the superspinar depends on a delicate balance
between the transmission coefficients through the ``inner'' and ``outer''
potential barriers.

A possible objection against instability could be the following.  For
sufficiently fast rotation (perhaps even for spins as low as $a\sim 6M$,
i.e. at the higher end of the viable range identified by
Ref.~\cite{Takahashi:2010pw}), if unstable modes exist in the eikonal limit
($l=m\to\infty$) their imaginary part will be small, even in the case ${\cal
  R}=1$.  The ergoregion instability is due to ergoregion modes leaking to
infinity through the potential barrier, but the tunneling becomes less and
less effective as the modes behave more and more like particles, because the
amplitude transmitted to infinity scales as $\exp(-L/\lambda)$ (where $L\sim
M$ is the width of the barrier and $\lambda$ is the mode's wavelength).
It is therefore conceivable that the imaginary part might be so tiny that
these modes can be considered stable for all practical purposes. 

However, accretion is known to spin superspinars {\em down}
\cite{ReinaTreves:1979}. According to Ref.~\cite{bardeen}, a BH which is
initially nonrotating gets spun up to the extremal limit $a=M$, where it
cannot be spun up any more \cite{bardeen,Wald:1974}, by accreting a mass
$\Delta M=(\sqrt{6}-1) M_{\rm in}=1.4495 M_{\rm in}$ ($M_{\rm in}$ being
the initial BH mass). This corresponds to the accretion of a gas mass $\Delta M_0=1.8464 M_{\rm in}$,
of which $\Delta M$ falls into the BH and $\Delta M_0-\Delta M$ is dissipated by the disk's viscosity
into electromagnetic radiation.
Similarly, a superspinar with $a/M=7$ gets spun down to
$a/M=1.5M$ (where the ergoregion instability is always effective) by accreting
 a mass $\Delta M=1.730 M_{\rm in}$ (corresponding to a gas mass $\Delta M_0=2.295M_{\rm in}$).
The two processes (spin-up of a Schwarzschild BH to the extremal Kerr limit, and
spin-down of a superspinar from $a/M=7$ to $a/M=1.5$) 
involve amounts of accreted material of the same order of magnitude, hence 
the corresponding timescales too will be comparable.  Supermassive BHs are expected to be spun up
to the extremal Kerr limit by coherent accretion%
\footnote{It has been proposed that supermassive BHs may accrete small lumps
  of material with essentially random orientations of the orbital angular
  momentum. This ``chaotic accretion'' results (on average) in a spin-down of
  the BH \cite{King:2008}, so it is very hard to produce fast spinning BHs at
  all (whereas spin estimates as large as $a=0.989^{+0.009}_{-0.002}$ have
  been reported \cite{Brenneman:2006hw}). Therefore it should be even harder
  to produce superspinars by chaotic accretion. Binary BH mergers are also
  known to always produce spins below the Kerr limit
  \cite{Sperhake:2009jz,Berti:2008af,2008ApJ...679.1422R,2008PhRvD..78d4002R,2009ApJ...704L..40B,Kesden:2010},
  so one would be left only with the possibility of postulating that
  supermassive superspinars are born in the early Universe due to high-energy
  physics effects beyond the realm of classical general relativity.}
on a timescale much smaller than the Hubble time \cite{Volonteri:2005}, so a
superspinar should be spun down to the unstable region on a timescale much
smaller than the Hubble time. For this reason, the existence of supermassive
superspinars is unlikely in the real Universe.

The situation is slightly different for stellar-mass superspinars. Analytical
arguments \cite{King:1999aq} and population synthesis calculations
\cite{Belczysnki:2008} show that BHs in binaries essentially retain the spin
they had at birth, so it is unclear whether accretion would be efficient
enough to destabilize a superspinar. On the other hand, as far as we know, no
realistic collapse scenario leading to the formation of stellar-mass
superspinars has been proposed so far. Typical equations of state lead to
compact stars rotating with $a/M\lesssim 0.7$ \cite{Cook:1993qr,Berti:2003nb}.
Polytropic
differentially rotating stars with $a\approx1.1M$ can in principle exist
\cite{GRS,whisky}, but they are stable. Even if depleted of 99\% of
their pressure and induced to collapse, these stars do not form a BH and
produce either a supermassive star (which will collapse to a BH with $a<M$ when
enough angular momentum has been shed in gravitational waves) or a
stable, rapidly rotating star.

For $r_0/M>0$ there is a second family of unstable modes with $m=0$ that
cannot be superradiant modes.  Figs.~\ref{fig:m012}~and~\ref{fig:m0_R0} show that,
for fixed values of $a/M$, these modes only exist in a limited range of
$r_0/M$. This range corresponds to the blue dot-dashed line (${\cal R}=1$
case) and to the solid black line (${\cal R}=0$ case) on the left of
Fig.~\ref{fig:rcrit}, showing that this family of unstable modes only exists
for $a/M\lesssim1.12$. 

These unstable modes are related to the existence of \emph{stable} ``polar''
null circular orbits, i.e. circular non-equatorial orbits with vanishing
azimuthal component of the orbital angular momentum ($L_z=0$) \cite{Dolan}.
Eq.~(\ref{a118}) of Appendix \ref{app:geod_Kerr} (see also
Fig.~\ref{fig:ang_mom}) shows that for the Kerr spacetime such orbits exist
when $1<a/M<\left(-3+2 \sqrt{3}\right) \approx1.17996$. For $l=4$ and ${\cal
  R}=1$ the instability range for modes with $m=0$ is
$1<a/M\lesssim1.14$. However the upper limit of this range is a slowly
increasing function of $l$, and it is plausible that in the eikonal limit it
should tend to $\left(-3+2 \sqrt{3}\right) \sim1.18$.

In conclusion, let us discuss the case $r_0/M<0$ (with ${\cal R}=1$),
summarized in Fig.~\ref{fig:dotti}.
Now the ring singularity at $r/M=0$ is naked, and the spacetime also possesses
closed timelike curves \cite{Carter:1968rr}.  Therefore it is not surprising
that an infinite number of unstable modes exist also at the linear level
\cite{Dotti:2008yr}. At variance with the ergoregion instability, in the
present case the imaginary part of the frequency (and therefore the
``strength'' of the instability) grows roughly quadratically with $a/M$
(cf. the right panel of Fig.~\ref{fig:dotti}). The same kind of instability
has been found in charged, spherically symmetric BHs with naked singularities
\cite{Dotti:2006gc} and therefore it is not related to rotation, but to
causality violation (see also the discussion at the end of
Ref.~\cite{ReinaTreves:1979}). As a matter of fact, we could not find any mode
belonging to this family when $r_0/M\geq0$, i.e. when the naked singularity is
covered.

In summary: superspinars are plagued by several instabilities for both
perfectly reflecting boundary conditions (${\cal R}=1$ at $r=r_0$) and
perfectly absorbing boundary conditions (${\cal R}=0$ at $r=r_0$). The
instability of modes with $l=m$ is related to superradiant scattering. When
$r_0\sim M$, unstable modes with $m=0$ also exist below some critical rotation
parameter: this instability is related to the existence of stable polar null
circular orbits in the spacetime (cf. Appendix \ref{app:geod_Kerr}). Finally,
when $r_0/M<0$, a third family of $m=0$ modes exists \cite{Dotti:2008yr}. This
third family of unstable modes is probably related to the existence of naked
singularity and closed timelike curves in the spacetime.

\section{Conclusions}

The results reported in this paper indicate that superspinars are unstable
independently of the boundary conditions imposed at the ``excision radius''
$r_0$ and in a significant region of the two-dimensional parameter space
$(a/M,\,r_0/M)$, if not in the whole parameter space.
The most effective instability at low rotation rates corresponds to the
$l=m=2$ (superradiant) mode, but when $a\sim M$ and $r_0\sim M$ unstable modes
with $m=0$ also exist.  The $l=m=2$ mode eventually becomes stable at large
rotation rates, but unstable modes with $l=m\gg1$ are expected to exist for
any value of $a/M$, at least for ${\cal R}=1$. While the instability timescale
of higher-$l$ modes may turn out to be very long, making them marginally
stable for practical purposes, the low-$l$ instability (which affects
superspinars with $a/M\lesssim2$) takes place on a dynamical
timescale. Accretion is known to spin superspinars
down~\cite{ReinaTreves:1979}, so our results indicate that superspinars are
unlikely astrophysical alternatives to Kerr BHs.

One possible objection is that, in order to assume ingoing boundary conditions
at the surface of the superspinar, we must assume that general relativity is
modified in that region.
Such a modification of general relativity in the excised, high-curvature
region surrounding the singularity is implicit in the original superspinar
proposal by Gimon and Horava~\cite{Gimon:2007ur}, who invoke string theory in
order to violate the Kerr bound $a\leq M$.  We stress, however, that our
results hold for a wide class of theories of gravity. Many proposed
alternative theories of gravity admit the Kerr spacetime as an exact
solution~\cite{psaltis,barausse_sotiriou}.  Among these theories, we focus on
the large class consisting of Brans-Dicke gravity (with or without a
potential), and theories that can be reduced to Brans-Dicke theory with a
potential via a conformal transformation (e.g. $f(R)$ gravity, both in the
metric and Palatini formalism~\cite{thomas_rev,wands}).  All of these theories
admit Kerr-(anti) de Sitter as an exact solution if the scalar field is
constant. When perturbed, these solutions satisfy different equations in
general relativity and in modified gravity
theories~\cite{barausse_sotiriou,will}, due to the presence of an extra scalar
degree of freedom (the Brans-Dicke scalar), so one might naively expect the
stability properties of the Kerr spacetime to be different. However, one can
redefine the tensor modes via a conformal transformation so that the
\textit{vacuum} tensor and scalar perturbation equations decouple at linear
order \cite{will}. Basically this happens because the Brans-Dicke action
reduces to the Einstein-Hilbert action plus a minimally coupled scalar field
in the Einstein frame, if no matter fields are present
\cite{wands}. Therefore, the tensor modes satisfy the same equations in
general relativity as in Brans-Dicke theory (or in any other theory that can
be recast in Brans-Dicke form via a conformal transformation.) This means that
Eqs.~\eqref{wave_eq} and \eqref{angularwaveeq}, which are the starting points
of our analysis, retain their validity, and therefore that the instability
operates in a wider class of gravity theories.

Our results do not imply that there cannot be stable ultracompact objects with
$J/M^2>1$. However, they do imply that either (i) Einstein's gravity should be
modified in such a way as to retain Kerr as a solution, while at the same time
allowing the tensor modes and the ``extra'' modes to couple at linear order,
or (ii) the structure of astrophysical superspinning objects, if they exist at
all, is not described by the simple Kerr-based superspinar proposal of
Ref.~\cite{Gimon:2007ur}.

\begin{acknowledgements}
It is a pleasure to thank Ted Jacobson for helpful discussions on the problems
discussed in this paper, as well as Cosimo Bambi for reading an early version
of the manuscript and providing useful comments and suggestions. We also thank
Sam Dolan for sharing with us some preliminary material on the relation
between geodesics and linear perturbations in the Kerr spacetime. Enrico
Barausse acknowledges support from NSF Grant PHY-0903631. Emanuele Berti's
research was supported by NSF grant PHY-0900735.  Vitor Cardoso acknowledges
financial support from Funda\c c\~ao Calouste Gulbenkian and from a
``Ci\^encia 2007'' research contract. This work was partially supported by FCT
- Portugal through projects PTDC/FIS/098025/2008, PTDC/FIS/098032/2008,
PTDC/CTE-AST/098034/2008, CERN/FP/109306/2009, CERN/FP/109290/2009.
\end{acknowledgements}


\appendix
\section{Geodesics in $D$-dimensional Kerr spacetimes}\label{app:geod_Kerr}
\begin{figure}[htb]
\begin{center}
\begin{tabular}{c}
\includegraphics[scale=0.31,clip=true]{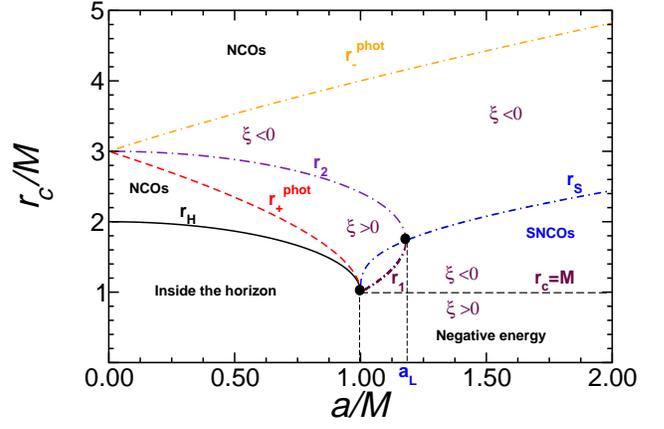}
\end{tabular}
\end{center}
\caption{\label{fig:ang_mom} The sign of $\xi=L_z/E$ as a function of $a/M$
  and the radius $r_c/M$ of null circular orbits. When $a>M$, $\xi\to\infty$
  at $r_c=M$. Stable circular null orbits exist when $a>M$ for $r_c<r_S$, and
  have negative energy when $r_c<M$. Regions marked with NCOs are those where
  no circular orbits exist. Orbits of constant radius $r_1$ are \emph{stable}
  ``polar'' null circular orbits.}
\end{figure}

The main goal of this appendix is to study the existence of stable null
circular orbits (SNCOs) with negative energies in Kerr spacetimes. We are
interested in these orbits because the very existence of SNCOs (or more
generally, the existence of stable null bound orbits) with \textit{negative}
energies is enough to show that a spacetime is subject to the ergoregion
instability, \textit{provided that} purely reflecting boundary conditions are
imposed at the excision surface $r=r_0$. For completeness we consider
$D$-dimensional Kerr spacetimes with only one nonzero angular momentum
parameter, and we specialize to the ``ordinary'' $D=4$ case at the end.  Our
main results for four-dimensional Kerr spacetimes are summarized in
Fig.~\ref{fig:ang_mom}. The meaning of the different curves on this plot is
explained below.

The metric of a $D$-dimensional Kerr BH with only one nonzero angular momentum
parameter is given in Boyer-Lindquist-type coordinates by \cite{Myers:1986un}
\begin{eqnarray}
ds^2&=&
-\frac{\Delta_D-a^2\sin^2\vartheta}{\Sigma}dt^2
-\frac{{2a(r^2+a^2-\Delta_D)\sin^2\vartheta}}{\Sigma}
dtd\varphi \nonumber\\
&&{}+\frac{(r^2+a^2)^2-\Delta_D a^2 
\sin^2\vartheta}{\Sigma}\sin^2\vartheta d\varphi^2+\frac{\Sigma}{\Delta_D}dr^2
\nn\\
&&{}+{\Sigma}d\vartheta^2+r^2\cos^2\vartheta d\Omega_{D-4}^2,
\label{metric}
\end{eqnarray}
where $\Sigma=r^2+a^2\cos^2\vartheta$, $\Delta_D=r^2+a^2-M_D r^{5-D}$,
$d\Omega_{D-4}^2$ denotes the metric of the unit $(D-4)$-sphere and $M_D$ and
$a$ are related to the physical mass $M$ and angular momentum $J$ of the
spacetime
\be
M_D=\frac{16\pi M}{D-2}A_D\,,\qquad a=\frac{D-2}{2}\frac{J}{M}\,,
\ee
with $A_D=(2\pi)^{(1-D)/2}\Gamma[(D-1)/2]$. 
The outer horizon is defined as the largest real root of $r_H^2+a^2-M_D
r_H^{5-D}=0$.
\subsection{Equatorial Null Geodesics}
For null geodesics in the equatorial plane ($\theta=\pi/2$) of the spacetime
(\ref{metric}), the radial geodesic equation reads
\be
E^{-2}\dot{r}^2=V_{\rm eff}=R(r)/r^4=1+\frac{M_D}{r^{D-1}}(\xi-a)^2-\frac{\xi^2-a^2}{r^2}\,,
\ee
where $\xi=L_z/E$ and where the dot denotes derivatives with respect to the
dimensionless affine parameter. Conditions for circular orbits are $V_{\rm
  eff}(r_c)=V'_{\rm eff}(r_c)=0$.  The condition $V_{\rm eff}(r_c)=0$ implies
\be
\xi=\frac{-a M_D \pm \sqrt{r_c^{2 (D-3)}\Delta_D(r_c)}}{r_c^{D-3}-M_D}
\ee
for direct and retrograde orbits, respectively. For $D=4$ the outer horizon is
located at $r=r_H=M+\sqrt{M^2-a^2}$ and (of course) the Kerr bound implies
$a/M\leq 1$. The condition $V'_{\rm eff}(r_c)=0$ then leads to three different
solutions:
\beq
r^{\rm phot}_{\pm}&=&2M
\left\{1+\cos\left[\frac{2}{3}
\cos^{-1}\left(\mp\frac{a}{M}\right)\right]\right\}\,,\nn\\
r_{c-}&=&2M-{\rm Re}[\beta]-\sqrt{3}{\rm Im}[\beta]\,,\label{rcs}
\eeq
%
where 
\be
\beta=\left[M\left(-M^2+2a^2+2a\sqrt{a^2-M^2}\right)\right]^{1/3}\,.
\ee
The three solutions are all real. Orbits with $r_c=r^{\rm phot}_{+}$
($r_c=r^{\rm phot}_{-}$) correspond to unstable direct (retrograde) circular
orbits lying outside the horizon, whereas $r_{c-}<r_H$ and therefore this
solution does not correspond to physical circular orbits.  For $a>M$ there is
only one real solution
\be
r^{\rm
  phot}_{-}=2M
\left\{1+\cosh\left[\frac{2}{3}\cosh^{-1}\left(\frac{a}{M}\right)\right]\right\}
\,,
\label{rcs2}
\ee
corresponding to an unstable null circular orbit. This is shown in
Fig.~\ref{fig:ang_mom}.  It is easy to show that the same qualitative results
hold also when $D\geq5$.  Therefore no stable null equatorial circular orbits exist in
$D$-dimensional Kerr spacetimes with a single angular momentum parameter.

\subsection{Non-equatorial null geodesics}
Let us now focus on non-equatorial null geodesics in $D$-dimensional Kerr
spacetimes with a single spin parameter. We shall follow and generalize the
approach discussed in Chandrasekhar's book \cite{Chandra}.

The separability of the Hamilton-Jacobi equation in Kerr spacetime was proved
by Carter, who also discovered an additional constant of motion $Q$ (the
``Carter constant'') besides the energy, the angular momentum and the norm of
the four-velocity \cite{Carter:1968rr,Carter:1968ks}. The same procedure can
be easily generalized to $D$-dimensional Kerr spacetimes with a single spin
parameter, given by  Eq.~(\ref{metric}).
Our basic equations are 
\begin{align}
&E^{-2}\Sigma^2\dot{r}^2=R(r)\label{R}\\
&=r^4+(a^2-\xi^2-\eta)r^2+M_D[\eta+(\xi-a)^2]r^{5-D}-a^2\eta\,,\nn
\\
&E^{-2}\Sigma^2\dot{\theta}^2=\Theta(\theta)\label{theta}\\
&=\eta+a^2\cos^2\theta-\xi^2\cot^2\theta\,,\nn\\
&E^{-1}\Delta_D\Sigma\,\dot{\varphi}= \Phi(r,\theta)\\
&=\xi \Delta_D  \csc^2\theta-a [a \xi+\Delta_D -(r^2+a^2)]\,,\nn\\
&E^{-1}\Delta_D\Sigma\,\dot{t}=T(r,\theta)\label{Vt}\\
&=(r^2+a^2)^2-a^2 \Delta_D  \sin ^2\theta+a \xi [\Delta_D -(r^2+a^2)]\,,\nn
\end{align}
where we use a dot to denote derivatives with respect to the dimensionless
affine parameter, and where $\xi=L_z/E$, $\eta=Q/E$ are two constants of
motion. Notice that the angular equation (\ref{theta}) does not depend on $D$,
and that the equations above reduce to Eqs.~(190) and (191) of
Ref.~\cite{Chandra} when $D=4$.

\subsubsection{Proof of no planar bounded orbits in $D$ dimensional Kerr spacetimes}
A relevant question is whether non-equatorial planar orbits exist in these
spacetimes. The conditions for a planar orbit ($\theta=\theta_0=$constant) are
$\Theta(\theta_0)=0=\Theta'(\theta_0)$. From Eq.~(\ref{theta}) we see that
these conditions are fulfilled on the equatorial plane ($\theta_0=\pi/2$) only
if $\eta=0$. For $\theta_0\neq\pi/2$ planar orbits exist if
\be
\eta=-a^2\cos^4\theta_0\,,\qquad \xi=\pm a\sin^2\theta_0\,.\label{shear-free}
\ee

For the ``plus'' branch of the solutions above, the radial equation (\ref{R})
simply becomes $\dot{r}=\pm E$. These geodesics are unbound and describe
shear-free null-congruences \cite{Chandra}. The ``minus'' branch of the
solutions describes non-equatorial planar orbits. These solutions only exist
when $D=4$ and $a<M$. Moreover they always lie inside the event horizon, and
therefore they do not correspond to physical orbits. For these reasons we do
not discuss them further.

\subsection{Non-equatorial, circular orbits}
Since no planar, non-equatorial circular orbits exist in $D-$dimensional Kerr
spacetimes, let us focus on non-equatorial, circular orbits, i.e. orbits with
constant radius but which are not planar (i.e. $\theta$ is not
constant). These orbits are periodic
\cite{Hughes:1999bq,Hughes:2001jr,Barausse:2007tk} and they are often called
``spherical orbits'' in the literature, but here we adopt the term ``circular
orbits'' as in Refs.~\cite{Hughes:1999bq,Hughes:2001jr,Barausse:2007tk}.

The conditions for null circular orbits, $R(r_c)=0$ and $R'(r_c)=0$, read
\begin{align}
&r_c^4+(a^2-\xi^2-\eta)r_c^2
\label{R1}
\\
&+M_D[\eta+(\xi-a)^2]r_c^{5-D}-a^2\eta=0\,, 
\nn
\\
&4r_c^3+2(a^2-\xi^2-\eta)r_c\label{R2}
\\
&+(5-D)M_D[\eta+(\xi-a)^2]r_c^{4-D}=0\,,
\nn
\end{align}
which can be solved for $\xi$ and $\eta$ as functions of $r_c$. There are two
sets of solutions:
\beq
\xi&=&\frac{r_c^2+a^2}{a}\,,\qquad\eta=-\frac{r_c^4}{a^2}\,, \label{sol1}
\eeq
and
\beq
\xi&=& \frac{a^2(D-5)M_D r_c^3+(D-1)M_D r_c^5-2
  r_c^D(a^2+r_c^2)}{a(D-5)M_Dr_c^3+2a r_c^D}\,,\nn\\
\eta&=&
\left[a(D-5)M_Dr_c^3+2a r_c^D\right]^{-2}\nn\\
&\times&\Bigl\{4M_D r_c^{5+D}(2a^2(D-3)+(D-1)r_c^2)\nn\\
&-&(D-1)^2M_D^2r_c^{10}-4r_c^{2(D+2)}\Bigr\} \,.
\label{sol2}
\eeq
The first set of solutions implies $\theta={\rm constant}$ and indeed reduces to the ``minus'' branch of solutions~(\ref{shear-free}), which do not correspond to physical orbits.

The second set of solutions, Eqs.~(\ref{sol2}), can describe bound orbits.
The condition of stability is simply $R''(r_c)<0$. By differentiating
Eq.~(\ref{R}) twice and using Eqs.~(\ref{sol2}) we obtain the following
expression for $R''(r_c)$:
\beq\label{stability}
R''(r_c)&=&\left[(D-5)M_D r_c^3+2 r_c^D\right]^{-2}\\
&\times&\Bigl\{-8(D-5)(D-1)M_D^2 r_c^8+32 r_c^{2(D+1)}\nn\\
&+&16(D-5) M_D  r_c^{D+3}\left[a^2(D-3)+(D-1)r_c^2\right]\Bigr\}\,.
\nn
\eeq
The stability of null circular orbits depends on the sign of the expression
above. It is possible to show that stable circular orbits exist for $D=4$ and
$a>M$, but not for $D\geq5$. Therefore in the following we will specialize to
$D=4$ spacetime dimensions.

\subsubsection{$D=4$ Kerr spacetime. }
When $D=4$, Eqs.~(\ref{sol2}) read
\beq
\xi&=&\frac{r_c^2(3M-r_c)-a^2(r_c+M)}{a(r_c-M)}\,,\label{sol2d4_xi} \\
\eta&=&\frac{r_c^3\left[4a^2M-r_c(r_c-3M)^2\right]}{a^2(r_c-M)^2}\,.
\label{sol2d4_eta}
\eeq
These equations correspond to Eqs.~(224) and (225) of Ref.~\cite{Chandra}, and
they can be used to define the shadow cast by Kerr BHs or superspinars
\cite{Bambi:2010hf,Hioki:2009na}. When $\eta=0$, from Eq.~\eqref{sol2d4_eta}
we have $4a^2M-r_c(r_c-3M)^2=0$, which defines the equatorial orbits
(\ref{rcs})-(\ref{rcs2}).  In general, however, the constant of motion $\eta$
can be positive or negative. When $\eta<0$, Eqs.~(\ref{sol2d4_xi})-\eqref{sol2d4_eta}, together with
Eq.~(\ref{theta}) for the $\theta$-motion, implies that orbits of constant
radius are not allowed \cite{Chandra}. When $\eta\geq0$ circular orbits are
allowed, and according to Eq.~(\ref{sol2d4_eta}) they must satisfy the condition
$4a^2M-r_c(r_c-3M)^2\geq0$. For $a<M$ this condition reads
\be
r^{\rm phot}_+<r_c<r^{\rm phot}_-\,,\nonumber
\ee
where $r^{\rm phot}_{-}$ and $r^{\rm phot}_+$ refer to retrograde and
direct unstable photon orbits in the equatorial plane [Eq.~\eqref{rcs}]. More importantly for
the analysis of superspinars, when $a>M$ the condition $\eta>0$ reads
\be
r_c<r^{\rm phot}_{-}\,,
\ee
where $r^{\rm phot}_{-}$ is given by Eq.~(\ref{rcs2}). Notice that the
condition above includes also the singular case $r_c=M$ (in fact
$\xi,\eta\to\infty$ when $a>M$ and $r_c\to M$).

When $D=4$, from Eq.~(\ref{stability}) we see that stable circular orbits
exist whenever the orbital radius $r_c$ satisfies the relation $r_c<r_S$, with
\cite{Hioki:2009na}
\begin{equation}\label{rs1}
\frac{r_S}{M}=\left\{
\begin{array}{ll}
\displaystyle 
1+\left[\left(\frac{a}{M}\right)^2-1\right]^{1/3}\mbox{   for } a>M\,,
\\
\displaystyle
1-\left[-\left(\frac{a}{M}\right)^2+1\right]^{1/3}\mbox{  for } a<M\,.
\end{array}
\right.
\end{equation}
Null circular orbits with radii smaller than this critical radius are
\emph{stable}. When $a>M$, $r_S<r^{\rm phot}_{-}$, i.e. stable circular orbits
are allowed.
When $a<M$ the critical radius $r_S$ is covered by the horizon, and it becomes
``visible'' to external observers only when $a>M$. Therefore {\em stable null
  circular orbits may exist only for $a>M$}, while orbits with $r<r_S$ around
BHs with $a<M$ do not have a physical meaning because they lie inside the
horizon.

By substituting Eq.~(\ref{rs1}) into Eqs.~(\ref{sol2d4_xi}) and~\eqref{sol2d4_eta} we can compute the
corresponding critical parameters $\eta(r_S)=\eta_S$ and $\xi(r_S)=\xi_S$:
\beq
\frac{\eta_S}{M^2}&=&\frac{3M^2}{a^2}\left(1+\left[\left(\frac{a}{M}\right)^2-1\right]^{1/3}\right)^4 \,,\nn\\ 
\frac{\xi_S}{M}&=&-\frac{a}{M}+\frac{3M}{a}\left(1-\left[\left(\frac{a}{M}\right)^2-1\right]^{2/3}\right)\,.\nn
\eeq
For a given value of $a/M$, when $\eta=\eta_S$ and $\xi=\xi_S$ we have a
marginally stable orbit. If instead $\xi\lesssim\xi_S$ we have a stable orbit,
while $\eta\lesssim\eta_S$ gives a stable orbit if $a>3M$ and
$\eta\gtrsim\eta_S$ gives a stable orbit if $a<3M$.  However, these are only
sufficient conditions, because other stable orbits may exist, far from the
critical values $\eta_S$ and $\xi_S$. In fact, depending on the value of the
spin we can have different situations: (i) for $a<3 M$, if $\eta<\eta_S(a)$
there is only one stable circular orbit (with $r_c<M$), while for
$\eta>\eta_S(a)$ we have two stable orbits: one with $r_c<M$ and one with
$r_c>M$; (ii) for $a>3 M$, when $\eta<27M^2$ we have only one stable circular
orbit (with $r_c<M$); when $27M^2<\eta<\eta_S(a)$ we have two stable orbits
with $r_c > M$ and one with $r_c < M$; when $\eta>\eta_S(a)$ we have one
stable circular orbit with $r_c<M$ and one with $r_c>M$.  This can be
understood by plotting $\eta$ as a function of $r$, with $0<r<r_S(a)$, for
various values of $a$.

Also, let us consider the sign of the impact parameter $\xi=L_z/E$.
A study of Eq.~(\ref{sol2d4_xi}) shows that
there is a critical spin
\be
a_L= \sqrt{3 \left(-3+2 \sqrt{3}\right)} M \approx1.17996 M\,,\label{a118}
\ee
such that:
\begin{itemize}
 \item if $a>a_L$, then $\xi>0$ for $r_c<M$ and $\xi<0$ for $r_c>M$. Notice
   that $\xi$ diverges if $r_c=M$.
 \item if $M<a<a_L$, then $\xi>0$ for $r_c<M$ and for $r_1(a)<r_c<r_2(a)$
   (with $r_1,r_2>M$), whereas $\xi<0$ elsewhere. Notice that $\xi$ diverges
   if $r_c=M$.
 \item if $a\leq M$ then $\xi>0$ for $r_+<r_c<r_2(a)$, where $r_+$ is the
   outer Kerr horizon and $r_2<3M$. $\xi<0$ for $r_c>r_2(a)$.
\end{itemize}
The situation for a four-dimensional Kerr spacetime is summarized in
Fig.~\ref{fig:ang_mom}. Orbits of radius $r_1$ and $r_2$ carry vanishing
angular momentum ($L_z=0$) and therefore are called ``polar'' null
orbits. Orbits of constant radius $r_2$ are unstable polar null orbits, while
orbits of constant radius $r_1$ are \emph{stable} polar null orbits, and they
exist for $M<a<a_L$.

A relevant question to ask is whether the  null circular orbits that we have
identified have positive or negative energy. The sign of the energy is
determined by imposing that the geodesics be future oriented, i.e.  that the
derivative $\dot{t}$ of the coordinate time with respect to the affine
parameter [given by Eq.~\eqref{Vt}] be positive. (This is because 
the hypersurfaces $t=$ const are spacelike for any $r>0$ if $a>M$, and for any $r>r_H$ if $a\leq M$.)
By substituting Eq.~\eqref{sol2d4_xi} into Eq.~\eqref{Vt}, we find that for the
non-equatorial null circular orbits that we have identified we have 
\be
\dot{t}=\frac{E}{\Sigma} 
\left[\frac{r_c^2 (r_c+3M)}{r_c - M} + a^2 \cos^2\theta\right]\,.
\label{tdot}
\ee
Because these orbits cross the equatorial plane (as they have $\eta>0$), we
can evaluate Eq.~\eqref{tdot} for $\theta=\pi/2$. The energy $E$ is a constant
of motion, so it cannot change sign along the trajectory. Then it is clear
that all orbits have $E>0$, with the exception of orbits with $r_c/M<1$,
which, as we have seen, exist only for $a/M>1$. Indeed, it possible to show
explicitly that orbits with $r_c/M<1$ in Kerr spacetimes with $a/M>1$ have negative
energy at all times. Using Eq.~\eqref{theta}, one obtains that such orbits
have polar angle $\theta$ oscillating between $\pi/2+\theta_c$ and
$\pi/2-\theta_c$, with
\begin{align}
&\cos^2\theta_c=\frac{2 r_c\sqrt{M\Delta\left(a^2M+r_c^2 (2 r_c-3M)\right)}-\rho}{a^2 (r_c-M)^2}\,,\nn\\&
\end{align}
where $\rho=r_c^4-3M^2r_c^2+2a^2Mr_c$. One can show that $\cos^2\theta_c<1$ for $a/M>1$ and $r_c/M<1$. Using
this expression in Eq.~\eqref{tdot} it is then possible to show that the energy
must be negative all along trajectories with $a/M>1$ and $r_c/M<1$. The region
where stable negative-energy orbits exist is shown in Fig.~\ref{fig:ang_mom}.

Finally, let us suppose we have a compact object
rotating with $a>M$. According to the cosmic censorship conjecture, the
singularity at $r/M=0$ must be excised by some horizon-like one-way membrane
or by a reflecting surface. It is then natural to ask the question of what the
compactness of the object can be if one wants to excise all SNCOs with negative energies. Because such orbits exist for any $r_c<M$, if $a>M$, 
the maximum allowed compactness turns out to be $\mu_{\max}={M}/{r}=1$. Because 
orbits with $r_c\lesssim M$ lie far away from the singularity at
$r/M=0$, this maximum compactness is not expected to be altered by high-energy corrections.

\bibliography{superspinars}

\end{document}